\documentclass [twocolumn, aps, showpacs] {revtex4}
\usepackage{graphicx}
\begin{document}
\draft
\title {Interaction and disorder in bilayer counterflow transport at filling factor one}

\author {E. Tutuc}
 \altaffiliation[Current address:] { I.B.M. TJ Watson Research Center, Yorktown Heights, NY 10598}
\author {M. Shayegan}
\affiliation{Department of Electrical Engineering, Princeton
University, Princeton, NJ 08544}
\date{\today}
\begin{abstract}
We study high mobility, interacting GaAs bilayer hole systems
exhibiting counterflow superfluid transport at total filling
factor $\nu=1$. As the density of the two layers is reduced,
making the bilayer more interacting, the counterflow Hall
resistivity ($\rho_{xy}$) decreases at a given temperature, while
the counterflow longitudinal resistivity ($\rho_{xx}$), which is
much larger than $\rho_{xy}$, hardly depends on density. On the
other hand, a small imbalance in the layer densities can result in
significant changes in $\rho_{xx}$ at $\nu=1$, while $\rho_{xy}$
remains vanishingly small. Our data suggest that the finite
$\rho_{xx}$ at $\nu=1$ is a result of mobile vortices in the
superfluid created by the ubiquitous disorder in this system.
\end{abstract}
 \maketitle

Interacting bilayer systems of two-dimensional carriers in the
limit of zero tunnelling can exhibit superfluidity \cite{wenzee}
in a peculiar "counterflow" transport configuration, where
currents of equal magnitude are passed in opposite directions in
the two layers. This phenomenon occurs when the bilayer is subject
to a perpendicular magnetic field ($B$) so that the total Landau
level filling factor, $\nu$, is 1 (layer filling factor 1/2) and
the inter-layer interaction is sufficiently strong to stabilize a
quantum Hall state (QHS) \cite{yang}. The physics of this QHS can
be understood by pairing the particles in the lowest Landau level
of one layer with the vacancies in the other layer, hence forming
neutral objects (excitons) which condense at the lowest
temperatures \cite{levi}. A similar superfluid was predicted to
occur in a closely spaced electron-hole bilayer at $B=0$
\cite{lozovik}. Key experimental evidence for the particle-vacancy
pair formation at $\nu=1$, and the ensuing condensation, has been
provided so far by two types of experiments. Inter-layer
tunnelling measurements have shown a much enhanced tunnelling
conductivity in the $\nu=1$ QHS \cite{spielman}, a behavior
reminiscent of a Josephson junction's. Most recently, counterflow
transport measurements have revealed that {\it both} the
longitudinal ($\rho_{xx}$) and Hall ($\rho_{xy}$) resistivities
vanish in the $\nu=1$ QHS in the limit of zero temperature
($T\rightarrow0$) \cite{kellogg,tutuc}. The vanishing of
$\rho_{xy}$ is especially important since it directly demonstrates
that the counterflow current is carried by {\it neutral}
particles, that is particle-vacancy pairs which have zero
electrical charge and therefore experience no Lorentz force.

An outstanding puzzle is what causes the dissipation in the
currently available samples and why experimentally there is no
finite critical temperature below which the counterflow
dissipation vanishes as the theory predicts \cite{yang}. Here we
study the counterflow transport in strongly interacting, high
mobility GaAs hole bilayers as a function of total density ($p$)
as well as layer density imbalance. Our data show that, at a given
temperature, the $\nu=1$ QHS counterflow $\rho_{xy}$ decreases
when $p$ is reduced to increase the inter-layer interaction. But
the counterflow $\rho_{xx}$, which is always much larger than
$\rho_{xy}$, barely depends on density. Furthermore, small changes
in the layer densities can substantially change the counterflow
$\rho_{xx}$, while $\rho_{xy}$ remains vanishingly small. The
vanishing $\rho_{xy}$ at low temperatures demonstrates that
inter-layer interaction is responsible for the particle-vacancy
pairing. The counterflow dissipation, signaled by the finite
$\rho_{xx}$ at finite temperature, has been attributed to the
existence of disorder-induced mobile vortices which move across
the superfluid current, much like in superfluid helium
\cite{levi,huse,sheng}. Our observed dependences of counterflow
$\rho_{xx}$ on total density and layer density imbalance are
consistent with this picture.

Our sample is a Si-modulation-doped GaAs double-layer hole system
grown on a GaAs (311)A substrate. It consists of two, 15nm wide,
GaAs quantum wells separated by an 8nm wide AlAs barrier. The top
and bottom barriers are Al$_{0.2}$Ga$_{0.8}$As layers. We used a
Hall bar geometry of $100\mu$m width, aligned along the
[01$\bar{1}$] crystal direction. The Hall bar mesa has two current
leads at each end, and three leads for measuring the longitudinal
and Hall voltages across the bar. Diffused InZn Ohmic contacts are
placed at the end of each lead. We use a combination of front and
back gates to selectively deplete one of the layers near each
contact, in order to realize independent contacts to each layer
\cite{eisenstein-apl}. As grown, the densities were
$p_{T}=3.1\times10^{10}$ cm$^{-2}$ and $p_{B}=3.8\times10^{10}$
cm$^{-2}$ for the top and bottom layers, respectively. The
mobility along [01$\bar{1}$] at these densities is approximately
32 m$^2/$Vs \cite{anisotropy}. Top and bottom gates were added on
the active area to control the layer densities. The measurements
were performed down to $T=30$mK, and using low-current
(0.5nA-1nA), low-frequency lock-in techniques.

\begin{figure*}
\centering
\includegraphics[scale=0.63]{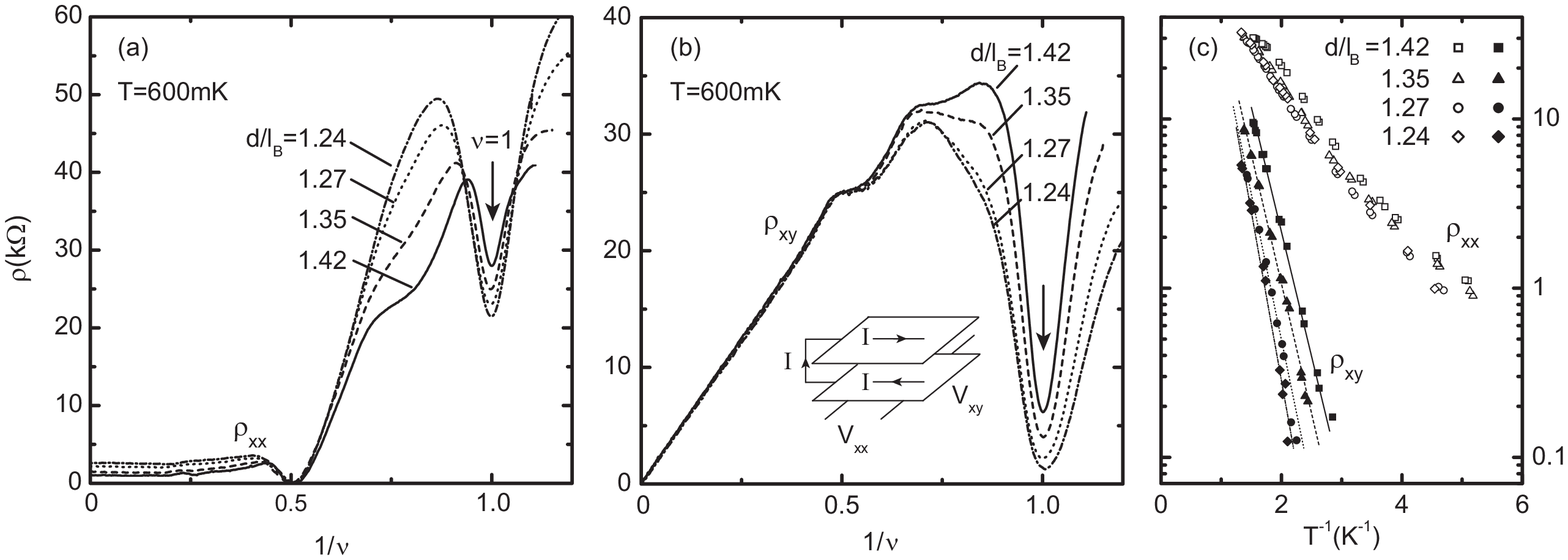}
\caption {\small{(a) $\rho_{xx}$ and (b) $\rho_{xy}$ counterflow
resistivities vs $1/\nu$ for different values of $d/l_B$,
corresponding to different total bilayer densities. (c)
Counterflow $\rho_{xx}$ (hollow symbols) and $\rho_{xy}$ (filled
symbols) at $\nu=1$ vs $T^{-1}$ for different $d/l_{B}$. The data
of all panels were taken when the two layers have equal densities.
Inset in (b) illustrates the geometry for counterflow
measurements.}}
\end{figure*}

In our counterflow measurements, two leads contacting opposite
layers at one end of the Hall bar are used to drive a current in
and out of the sample, while the leads at the other end are
shorted so that the same current, but in opposite directions,
flows in both layers [Fig. 1(b) inset]. We define the resistances
in the counterflow configuration as the corresponding voltage
drops along or across the Hall bar measured in one {\it single}
layer, the bottom layer in our case, divided by the current
flowing in each layer. The longitudinal resistivity $\rho_{xx}$ is
the measured longitudinal resistance divided by two, the number of
squares between the voltage probes in our sample, while the Hall
resistivity $\rho_{xy}$ is simply the Hall resistance. These
resistivities revert to those of a single layer when the coupling
between the two layers is negligible. In our counterflow
measurements a small fraction of the current injected in one layer
leaks unintentionally into the opposite layer. The leakage current
at $\nu=1$ is about 1$\%$ of the total current at 30mK and
increases with $T$ to 15$\%$ at 700mK; this behavior suggests that
the leakage is determined by factors (e.g. defects in the AlAs
barrier) other than the enhanced inter-layer tunnelling at $\nu=1$
\cite{spielman}. The leakage translates in a slightly lower
counterflow current, resulting in a small error in $\rho_{xx}$ and
$\rho_{xy}$. We emphasize that none of our conclusions are
affected by such errors.

The phase-space of the $\nu=1$ QHS is parametrized by the ratio
$d/l_{B}$, of the mean inter-layer spacing ($d$) and the magnetic
length ($l_{B}=\sqrt{\hbar/eB}$) at $\nu=1$. This parameter
quantifies the ratio between the intra- and inter-layer
interaction: when $d/l_{B}\gg1$ the latter is negligible and the
physics of the bilayer is essentially that of two independent
layers, while for $d/l_{B}\simeq1$ the inter-layer interaction is
sufficiently strong to give rise to collective bilayer phenomena
such as the $\nu=1$ QHS \cite{murphy}. Exploring this desired
regime of small values of $d/l_{B}$ poses experimental challenges.
Small $d$ and/or large $l_{B}$, hence low densities, are needed to
reduce $d/l_{B}$. Reducing $d$, however, increases the single
particle inter-layer tunnelling, which impedes the fabrication of
independent contacts and also can reduce the bilayer physics to
that of a single layer. From this perspective, GaAs {\it holes}
are desirable for exploring interacting bilayer phenomena: owing
to their relatively large effective mass, the inter-layer barrier
thickness can be reduced without substantially increasing the
tunnelling, hence placing the hole bilayers in a very strong
interacting regime with small tunnelling \cite{tunnelling}. Indeed
the data presented here correspond to the smallest $d/l_B$ at
$\nu=1$ yet reported for samples with independent contacts, also
evinced by the fully developed $\nu=1$ and vanishingly small
counterflow $\rho_{xy}$ at relatively high temperatures.

In Fig. 1 we present the counterflow $\rho_{xx}$ and $\rho_{xy}$
data when the two layers have equal densities (balanced). The
different $d/l_B$ indicated in the figure (1.24 to 1.42)
correspond to different $p$ ($4.5$ to $6.1\times10^{10}$
cm$^{-2}$). Data of Fig. 1(a,b) show $\rho_{xx}$ and $\rho_{xy}$
vs $1/\nu$, measured at $T=600$mK, illustrating that
$\rho_{xx}\gg\rho_{xy}$ at this temperature for all values of
$d/l_B$ explored here. The data of Fig. 1(b) further reveal that
$\rho_{xy}$ at $\nu=1$ decreases when $d/l_{B}$ is reduced. For
example, as shown in Fig. 1(b), at $T=600$mK $\rho_{xy}$ at
$\nu=1$ drops by more than a factor of 4 when $d/l_{B}$ is reduced
from 1.42 to 1.24. This observation demonstrates that the
inter-layer interaction is responsible for the particle-vacancy
pairing, signaled by the vanishing counterflow $\rho_{xy}$
\cite{footnote2}. Remarkably, the same change in $d/l_{B}$ induces
a much smaller change ($\simeq30\%$) in counterflow $\rho_{xx}$ at
$\nu=1$, as shown in Fig. 1(a).

Figure 1(c) summarizes the counterflow $\rho_{xx}$ and $\rho_{xy}$
data at $\nu=1$ vs $T^{-1}$, for various values of $d/l_{B}$.
These data further demonstrate that, in contrast to $\rho_{xy}$,
the counterflow $\rho_{xx}$ is much less dependent on $d/l_{B}$ at
any temperature. Fitting an exponential dependence
$\rho_{xy}\propto\exp(-\Delta_H/2T)$ to the data of Fig. 1(c)
yields a Hall energy gap $\Delta_{H}$ which increases from 6.8K to
9.2K, as $d/l_{B}$ is reduced from 1.42 to 1.24. The gap extracted
by fitting $\rho_{xx}\propto\exp(-\Delta/2T)$ to the $\rho_{xx}$
data is $\Delta\cong2K$, almost independent of $d/l_{B}$. This
finding is very revealing: unlike the pairing, which becomes
stronger when the bilayer is made more interacting (smaller
$d/l_{B}$), the dissipation is barely dependent on the inter-layer
interaction in our sample. We will discuss this observation in our
closing paragraph.

In view of the large counterflow $\rho_{xx}$ and $\rho_{xy}$
anisotropy at $\nu=1$, it is useful to examine the role of the
mobility anisotropy in our sample. Our measurements show that the
Hall drag remains close to the quantized value of $h/e^2$ as $T$
is increased, and hardly depends on the crystal direction along
which the current flows. The close quantization of the Hall drag
at higher $T$ is consistent with a small counterflow $\rho_{xy}$
\cite{tutuc} and suggests that the counterflow $\rho_{xy}$ does
not depend on the crystal direction. On the other hand, our
bilayer measurements on samples without independent contacts
(parallel flow) show that $\rho_{xx}$ at $\nu=1$ is about 2.5
times larger for current oriented along the low mobility
direction, and exhibits activated $T$ dependence with an energy
gap that is independent of the current direction. These findings
suggest that, while the counterflow $\rho_{xx}$ and $\rho_{xy}$
anisotropy at $\nu=1$ may be partially enhanced by the mobility
anisotropy in our samples, the latter is not the main factor
behind the observed counterflow anisotropy and the strong pairing.

Two noteworthy aspects of Fig. 1 data are (1) the large difference
between the counterflow $\rho_{xx}$ and $\rho_{xy}$, and (2) the
absence of a finite temperature below which the counterflow
transport is dissipationless. These can both be understood by
considering the motion of unpaired, mobile vortices in the
superfluid flow. Vortices in a superfluid are subject to a Magnus
force, perpendicular to the direction of the superfluid current,
which causes them to move {\it across} the superflow. The unpaired
vortex motion results in phase slip and implicitly, dissipation,
hence the finite values of $\rho_{xx}$ while $\rho_{xy}$ remains
close to 0. In the case of the $\nu=1$ QHS counterflow superfluid
the picture is slightly complicated because the vortices possess
electrical charge and dipole moment \cite{yang}, but the
conclusions of the above argument remain valid \cite{huse}.
Theoretical studies \cite{sheng,fertig} have further shown that
the source of unpaired vortices in the $\nu=1$ superfluid is
disorder. The experimentally observed absence of a critical
temperature below which the superfluid is dissipationless suggests
that mobile vortices are present at any finite $T$ in our sample.
As the temperature is lowered the vortices become pinned by
disorder, form a vortex glass, and lead to dissipationless flow
only in the $T=0$ limit. We emphasize, however, that a
quantitative understanding of the role of disorder, vortex
formation, and pinning in $\nu=1$ counterflow superfluid is
currently lacking.

\begin{figure}
\centering
\includegraphics[scale=0.28]{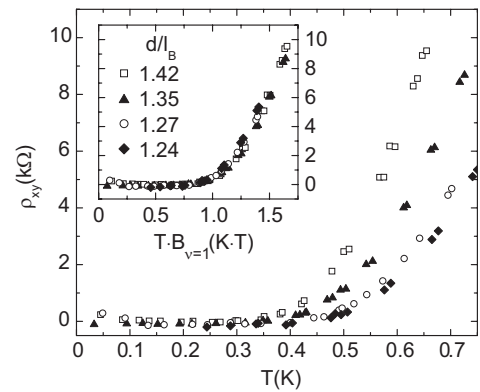}
\caption {\small{Counterflow $\rho_{xy}$ at $\nu=1$ for different
$d/l_B$, vs $T$. The inset shows an empirical scaling: the
$\rho_{xy}$ data of the main panel collapse approximately into a
single curve when plotted vs $T\cdot B_{\nu=1}$, where $B_{\nu=1}$
is the magnetic field at $\nu=1$.}}
\end{figure}

\begin{figure*}
\centering
\includegraphics[scale=0.61]{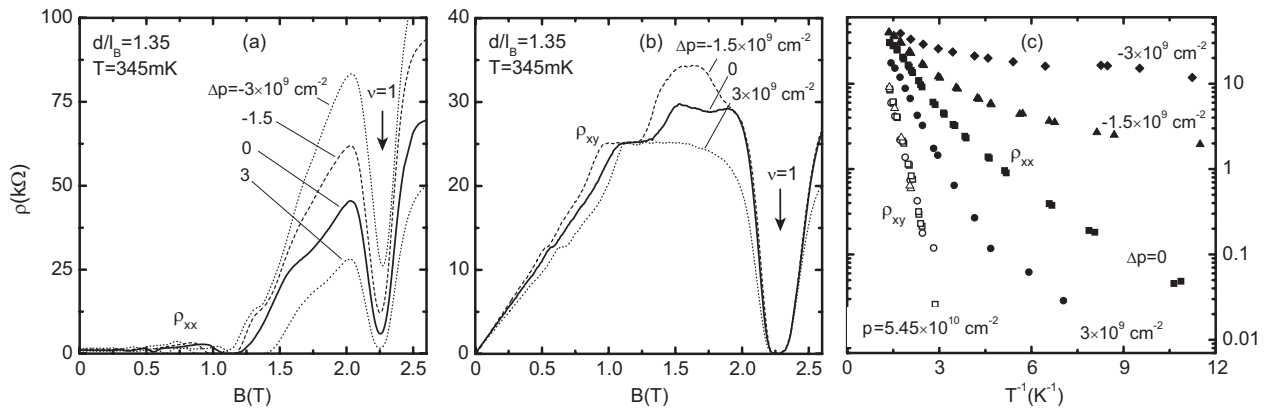}
\caption {\small{(a,b) Counterflow $\rho_{xx}$ and $\rho_{xy}$ vs
$B$ traces, measured at a constant total density
$p=5.45\times10^{10}$ cm$^{-2}$ ($d/l_B=1.35$) but for different
layer charge imbalance configurations, $\Delta p$. (c) Counterflow
$\rho_{xx}$ and $\rho_{xy}$ vs $T^{-1}$ at different values of
$\Delta p$, indicated by the different symbol shapes. Filled
(hollow) symbols represent $\rho_{xx}$ ($\rho_{xy}$).}}
\end{figure*}

It is instructive to compare our data with similar counterflow
measurements reported for GaAs {\it electron} bilayer systems
\cite{kellogg,wiersma}. In Ref. \cite{kellogg} the $T$ dependence
of $\rho_{xx}$ and $\rho_{xy}$ at $d/l_B=1.48$ was measured in the
range of 30-500mK and, in contrast to our GaAs hole data, little
difference was reported between counterflow $\rho_{xx}$ and
$\rho_{xy}$ in this $T$ range \cite{dlb}. On the other hand, in
Ref. \cite{wiersma}, the counterflow $\rho_{xy}$ is reported to be
much smaller than $\rho_{xx}$ at 40mK for a bilayer electron
system with a relatively large $d/l_B=1.57$, qualitatively
consistent with our hole data. While we cannot explain these
similarities and differences, we mention three distinguishing
factors of holes, besides their mobility anisotropy. First, the
larger cyclotron effective mass of GaAs holes vs electrons causes
more Landau level mixing; whether this leads to a stronger pairing
is not clear. Second, there is less inter-layer tunnelling in the
hole system \cite{tunnelling}. Third, it is possible that the
holes are more fully spin polarized at $\nu=1$ thereby increasing
the strength of the $\nu=1$ QHS \cite{spileman-nmr}.

It is interesting to examine the counterflow $\rho_{xy}$ data at
$\nu=1$ on a linear scale and down to the lowest $T$, as shown in
Fig. 2. For all $d/l_B$, the counterflow $\rho_{xy}$ remains small
up to a certain temperature at which it starts to increase
relatively sharply. As apparent from Fig. 2, this onset increases
as $d/l_B$ is reduced. In Fig. 2 inset we show an empirical
scaling of the $\rho_{xy}$ data at different values of $d/l_B$.
When the counterflow $\rho_{xy}$ is plotted vs $T\cdot B_{\nu=1}$,
where $B_{\nu=1}$ is the magnetic field at $\nu=1$, the data
points collapse onto a single curve. Although the decrease of
$\rho_{xy}$ with decreasing $d/l_{B}$ and the resulting stronger
pairing is expected, the origin of such simple scaling is unclear.
When extrapolated to large values of $d/l_{B}$ the scaling shown
of Fig. 2 must break down, since it does not predict a critical
$d/l_{B}$ for the disappearance of the $\nu=1$ QHS. Experimentally
such critical $d/l_B$ exists \cite{murphy} and in GaAs hole
bilayers with negligible tunnelling is between 1.6 and 1.9.

We now turn to a study of counterflow transport at $\nu=1$ as a
function of layer density imbalance. Using the top and bottom
gates on the active area, we keep $p$ constant while transferring
charge from one layer to another. We define the layer density
imbalance as $\Delta p=(p_B-p_T)/2$, and measure the counterflow
resistivites of the bottom layer. Depending on the sign of $\Delta
p$, the latter is either the majority density layer ($\Delta
p>0$), or the minority layer ($\Delta p<0$). In Fig. 3(a,b) we
show examples of counterflow $\rho_{xx}$ and $\rho_{xy}$ vs $B$
traces, measured at $T=345$mK for different values of $\Delta p$,
at constant $p=5.45\times10^{10}$ cm$^{-2}$ corresponding to
$d/l_B=1.35$. The data of Fig. 3(a) clearly demonstrate that
$\rho_{xx}$ at $\nu=1$ and therefore the dissipation, can be
greatly changed by a small layer density imbalance: a $\Delta p/p$
of $\simeq\%5$ changes the counterflow $\rho_{xx}$ by more than
one order of magnitude. A similar dependence of individual layer
$\rho_{xx}$ with density imbalance was recently reported in GaAs
electron bilayers \cite{wiersma}. More importantly, Fig. 3(b)
further reveal that the counterflow $\rho_{xy}$ at $\nu=1$ remains
vanishingly small at this $T$ for all values of $\Delta p$. The
data of Fig. 3(a,b) illustrate that the dissipation in counterflow
transport at $\nu=1$ dramatically changes when the layers are
imbalanced while the particle-vacancy pairing remains strong, as
evinced by the vanishing counterflow $\rho_{xy}$ at $\nu=1$. In
Fig. 3(c) we summarize our counterflow $\rho_{xx}$ and $\rho_{xy}$
vs $T^{-1}$ data, taken at different values of $\Delta p$ and at
constant $p=5.45\times10^{10}$ cm$^{-2}$. The data substantiate
the findings of Fig. 3(a,b), namely that the counterflow
$\rho_{xx}$ is very sensitive to layer density imbalance, while
the counterflow $\rho_{xy}$ is not.

Why does a rather small change in layer density distribution so
dramatically change the counterflow $\rho_{xx}$ at $\nu=1$, while
$\rho_{xy}$ remains vanishingly small? It is unlikely that this
behavior results from the small changes in the intra-layer
interaction energies of the two layers. Disorder, on the other
hand, can be the culprit. When $\Delta p<0$, the (bottom) layer,
which we probe, has a smaller density than in the balanced case
and is thus more prone to disorder since, e.g., the screening of
the ionized impurity potential is less effective. It is therefore
not surprising that its $\rho_{xx}$ at $\nu=1$ is larger than in
the balanced case. The converse is true when $\Delta p>0$. The
data of Fig. 3 also provide a natural clue for understanding the
results of Fig. 1. Lowering the density and therefore $d/l_B$ in
the balanced case has two consequences which influence $\rho_{xx}$
in opposite directions. First, it strengthens the inter-layer
interaction and leads to a lowering of $\rho_{xy}$, as observed.
Theoretically, we would also expect a reduction in $\rho_{xx}$
with decreasing $d/l_B$. But a second consequence of lowering the
density is to enhance the effective disorder in the bilayer system
and thus increase $\rho_{xx}$. We believe these two, compensating
effects are responsible for $\rho_{xx}$ in Fig. 1 being
essentially independent of $d/l_B$ while $\rho_{xy}$ significantly
decreases with decreasing $d/l_B$. The combination of the data of
Figs. 1 and 3 therefore suggests that the sample disorder, and the
ensuing mobile vortices \cite{sheng,huse,fertig}, are the likely
culprits for the finite counterflow dissipation in our samples.

We thank D.A. Huse, K. Yang, N.P. Ong, N. Bishop, R. Winkler and
O. Gunawan for discussions and DOE and NSF for support.


\begin{thebibliography}{10}
\bibitem{anisotropy}
The mobility of GaAs 2D hole systems grown on (311)A substrates is
lower along the [01$\bar{1}$] direction.
\bibitem{tunnelling}
The single particle inter-layer tunnelling energy for the sample
explored here is about 1$\mu$K.
\bibitem{footnote2}
Qualitatively similar data were reported in [5] for bilayer
electrons at higher $d/l_B$ (between 2.29 and 1.48) and a much
lower temperature ($T=50$ mK).
\bibitem{dlb}
Caution must be excercised when comparing various bilayer data
sets uniquely in terms of $d/l_B$. For example, growth
uncertainties in our sample prevent us to determine the absolute
value of $d$ to better than $\pm15\%$.
\end{thebibliography}
\end{document}